\documentclass{elsart5p}

\usepackage[usenames,dvipsnames]{color}
\usepackage{graphicx}
\usepackage{amsmath}
\usepackage{amssymb}

\newcommand{\be}{\begin{equation}}
\newcommand{\ee}{\end{equation}}
\newcommand{\K}{{\cal K}}
\DeclareMathOperator{\re}{Re}
\DeclareMathOperator{\myspan}{span}
\DeclareMathOperator{\sign}{sgn}
\DeclareMathOperator{\diag}{diag}
\DeclareMathOperator{\polar}{pol}
\newcommand{\MathC}{\mathbb{C}}

\begin{document}

\begin{frontmatter}

\title{An iterative method to compute the sign function of a
  non-Hermitian matrix and its application to the overlap Dirac operator at nonzero chemical potential}

\author[Regensburg]{J. Bloch},
\author[Wuppertal]{A. Frommer},
\author[Wuppertal]{B. Lang}, and
\author[Regensburg]{T. Wettig}
\address[Regensburg]{Institute for Theoretical Physics, University of
  Regensburg, 93040 Regensburg, Germany}
\address[Wuppertal]{Department of Mathematics, University of Wuppertal,
  42097 Wuppertal, Germany}
\date{26 April 2007}
\begin{abstract}
The overlap Dirac operator in lattice QCD requires the computation of
the sign function of a matrix.
While this matrix is usually Hermitian, it becomes non-Hermitian in the
presence of a quark chemical potential.
We show how the action of the sign function of a non-Hermitian matrix
on an arbitrary vector can be computed efficiently on large lattices
by an iterative method.
A Krylov subspace approximation based on the Arnoldi algorithm is
described for the evaluation of a generic matrix function.
The efficiency of the method is spoiled when the matrix has eigenvalues
close to a function discontinuity. 
This is cured by adding a small number of critical eigenvectors to the Krylov subspace,
for which we propose two different deflation schemes.
The ensuing modified Arnoldi method is then applied to the sign function, which
has a discontinuity along the imaginary axis.
The numerical results clearly show the improved efficiency of the method.
Our modification is particularly effective when the action of the sign
function of the same matrix has to be computed many times on different
vectors, e.g., if the 
overlap Dirac operator is inverted using an iterative method.

\end{abstract}

\end{frontmatter}

\section{Introduction}

The only systematic nonperturbative approach to quantum chromodynamics
(QCD) is the numerical simulation of the theory on a finite space-time
lattice.
For a long time, the implementation of chiral symmetry on the lattice
posed serious problems \cite{NN}, but these problems have recently been
solved in a number of complementary ways.
Perhaps the most prominent solution is the overlap Dirac operator
\cite{overlap}
which provides an exact solution of the Ginsparg-Wilson relation
\cite{Ginsparg:1981bj}.
However, the price one has to pay for this solution is the numerical
computation of the sign function of a sparse matrix $A$ of dimension $N$.
Here and in the following, \emph{computing some function $f$ of a
matrix $A$} is a short-hand for
\emph{computing $f( A ) \cdot x$,} where $x \in \MathC^{N}$,
i.e., determining the action of $f( A )$ on the vector $x$.
Typically $A$ is Hermitian, and efficient methods to compute its
sign function have been developed for this case
\cite{Neuberger:1998my,vandenEshof:2002ms}.

The phase diagram of QCD is currently being explored experimentally in
relativistic heavy ion collisions and theoretically in lattice
simulations and model calculations \cite{Misha06}.
To describe QCD at nonzero density, a quark chemical potential is
introduced in the QCD Lagrangian.
If this chemical potential is implemented in the overlap operator
\cite{Bloch:2006cd}, the matrix $A$ loses its Hermiticity,
and one is faced with the problem of computing the sign function of a
non-Hermitian sparse matrix.
On a small lattice, this can be done by performing a full
diagonalization and using the spectral matrix function definition
(see Eq.~\eqref{fA} below), but on
larger lattices one needs to resort to iterative methods to keep the
computation time and memory requirements under control.

The purpose of this paper is to introduce such an iterative method.
In the next section we describe the non-Hermitian problem in more detail
and briefly discuss the sign function for non-Hermitian matrices.
In Sec.~\ref{NumApp} we propose an Arnoldi-based method to make a
Krylov subspace approximation of a generic matrix function. 
The efficiency of this method is poor when computing the sign function of a
matrix having eigenvalues with small absolute real parts.
This is caused by the discontinuity of the sign function along the
imaginary axis.
In Sec.~\ref{Deflation} we enhance the Arnoldi method by
taking into account exact information about these critical eigenvalues. 
We use the method to compute the sign function occurring in the overlap Dirac operator of lattice QCD, and in Sec.~\ref{Results} we discuss the results obtained for two different lattice sizes.

\section{The overlap operator and the sign function}
\label{sec:sign}

The overlap formulation of the Dirac operator is a rigorous method
to preserve chiral symmetry at finite lattice spacing in a
vector-like gauge theory.  Its construction is based on successive
developments described in 
seminal papers by Kaplan, Shamir, Furman, Narayanan and Neuberger
\cite{Kaplan:1992bt,Shamir:1993zy,Furman:1994ky,overlap}.
In the presence of a non-zero quark chemical potential $\mu$, the massless
overlap Dirac operator is given by \cite{Bloch:2006cd}
\be 
D_{\text{ov}}(\mu) = 1 + \gamma_5\sign(H_\text{w}(\mu)) \:,
\label{Dov}
\ee
where $\sign$ stands for the sign function, $H_\text{w}(\mu)=\gamma_5 D_\text{w}(\mu)$, $D_\text{w}(\mu)$ is the Wilson-Dirac operator
at nonzero chemical potential
\cite{Hasenfratz,Kogut:1983ia} with negative
Wilson mass $m_\text{w} \in (-2,0)$,
$\gamma_5=\gamma_1\gamma_2\gamma_3\gamma_4$, and
$\gamma_\nu$ with $\nu=1,\ldots,4$ are the Dirac gamma matrices in
Euclidean space.
The Wilson-Dirac operator is a discretized version of the continuum
Dirac operator that avoids the replication of fermion species which
occurs when a naive discretization of the derivative operator is used.
It is given by
\begin{align}
\label{Dw}
[D_\text{w}(\mu)]_{nm} &= 
\delta_{n,m} \\
&  \hspace{-13mm}
   - \kappa \sum_{j=1}^3  (1+\gamma_j) U_{n,j} \delta_{n+\hat j,m} 
   - \kappa \sum_{j=1}^3 (1-\gamma_j)
        U^\dagger_{n-\hat j,j} \delta_{n-\hat j,m} \notag\\
& \hspace{-13mm}
   - \kappa  (1+\gamma_4) e^\mu U_{n,4} \delta_{n+\hat 4,m} 
   - \kappa (1-\gamma_4) e^{-\mu}
        U^\dagger_{n-\hat 4,4} \delta_{n-\hat 4,m} \:, \notag
\end{align}
where $\kappa = 1/(8+2m_\text{w})$ and $U_{n,\nu}$ is the $SU(3)$-matrix
associated with the link connecting the lattice site $n$ to $n+\hat\nu$.
The exponential factors $e^{\pm\mu}$ are responsible for the
non-Hermiticity of the operator. 
The quark field at each lattice site corresponds to 12
variables: 3 $SU(3)$ color components $\times$ 4 Dirac spinor
components.

For $\mu \ne 0$ the argument $H_\text{w}(\mu)$ of the sign function becomes
non-Hermitian, and we need to define the sign function for this case.
Consider first a given matrix $A$ with no particular symmetry
properties and a function $f$. Let $\Gamma$ be a 
collection of closed contours in $\MathC$ such that $f$ is analytic inside
and on $\Gamma$ and such that $\Gamma$ encloses the spectrum of $A$.
Then the function $f( A )$ of the matrix $A$ can be defined by \cite{Dunford}
\be
f(A) = \frac{1}{2\pi i} \oint_\Gamma f(z) (z I - A)^{-1} dz \:,
\label{fcontour}
\ee
where the integral is defined component-wise and $I$ denotes the
identity matrix.

From this definition it is easy to derive a spectral function
definition, even if the matrix is non-Hermitian.
If the matrix $A$ is diagonalizable, i.e., $A=U \Lambda U^{-1}$ with a diagonal
eigenvalue matrix $\Lambda=\diag(\lambda_i)$ and
$U\in\text{Gl}(N,\mathbb C)$, then
\begin{align}
  \label{fA}
  f(A) = U f(\Lambda) U^{-1}
  \intertext{with}
  f(\Lambda) = \diag(f(\lambda_i)) \:.
\end{align}
If $A$ cannot be diagonalized, a more general spectral definition of $f(A)$ can be
derived from Eq.~\eqref{fcontour} using the Jordan decomposition $A=U (\bigoplus_i J_i) U^{-1}$ with Jordan blocks
\be
J_i = 
\begin{pmatrix}
\lambda_i & 1  & \cdots &  0 \\
0 & \lambda_i  & \ddots & \vdots \\
\vdots & \ddots & \ddots &  1 \\
0 & \cdots & 0 & \lambda_i
\end{pmatrix} \:.
\ee 
Then, 
\be
f(A)= U \left(\bigoplus_i f(J_i)\right) U^{-1} \:,
\label{fAJordan}
\ee
where
\be
f(J_i) = 
\begin{pmatrix}
f(\lambda_i) & f^{(1)}(\lambda_i)  & \cdots & \frac{f^{(m_i-1)}(\lambda_i)}{(m_i-1)!} \\
0 & f(\lambda_i) & \ddots & \vdots \\
\vdots & \ddots & \ddots &  f^{(1)}(\lambda_i) \\
0 & \cdots & 0 & f(\lambda_i)
\end{pmatrix} 
\ee 
with $m_i$ the size of the Jordan block, see \cite{Golub}.  The superscripts denote derivatives of the function with respect to its argument.

Non-Hermitian matrices typically have complex eigenvalues, and applying
Eq.~\eqref{fA} or \eqref{fAJordan} to the sign function in Eq.~\eqref{Dov} requires the
evaluation of the sign of a complex number. 
The sign function needs to satisfy $[\sign(z)]^2=1$ and, for real $x$,
$\sign(x) = \pm 1$ if $x \gtrless 0$.
These properties are satisfied if one defines 
\be
\sign(z) \equiv \frac{z}{\sqrt{z^2}} = \sign(\re(z)) \:,
\label{sgnz}
\ee
where in the last equality the cut of the square root is chosen along the negative real axis. 
Using the definition \eqref{sgnz} in the spectral definition \eqref{fAJordan}
yields a matrix sign function in agreement with that used in \cite{Denman76,Rob80,Kenney91,Hig94}. Indeed, based on the general Jordan decomposition 
\begin{equation}
A = U \left( \begin{array}{cc} J_+ & \\ & J_- \end{array} \right) U^{-1} \:,
\end{equation}
where 
$J_+$ represents the Jordan blocks corresponding to the eigenvalues
with positive real part and $J_-$ those corresponding to the eigenvalues
with negative real part,
the spectral definition \eqref{fAJordan}
for the sign function becomes
\begin{equation} \label{signA}
\sign(A) = U \left( \begin{array}{cc} +I & \\ & -I \end{array} \right) U^{-1} \:,
\end{equation}
see also \cite{HoJo}.  This definition agrees with the result one
obtains when deriving Eq.~\eqref{Dov} from the domain-wall fermion
formalism at $\mu\ne0$ in the limit in which the extent of the fifth
dimension goes to infinity \cite{Bloch:2007xi}.

For any square matrix $A$ we have $\sign(A)^2 = I$, and a short
calculation \cite{Bloch:2006cd} shows that for this reason the overlap
operator $D_{\text{ov}}(\mu)$ as defined in Eq.~(\ref{Dov}) satisfies
the Ginsparg-Wilson relation
\begin{equation} \label{GW:eq}
\{D_{\text{ov}},\gamma_5\} = D_{\text{ov}} \gamma_5 D_{\text{ov}}\:.
\end{equation}
If $A$ is Hermitian, the polar factor $\polar(A) = A (A^\dagger
A)^{-1/2}$ of $A$ coincides with $\sign(A)$, and this fact has been
used successfully to develop efficient iterative methods for computing
the action of the matrix sign function on a vector
\cite{Borici:lanczos}. However, if $A$ is non-Hermitian, then in
general $\sign(A)\ne\polar(A)$ and $\polar(A)^2\ne I$.  Thus, for $\mu
\ne 0$, replacing $\sign(H_\text{w})$ by $\polar(H_{\text{w}})$ in the
definition of the overlap operator in Eq.~(\ref{Dov}) not only changes
the operator but also violates the Ginsparg-Wilson relation, as can also 
be seen in numerical experiments.  We conclude that the definition given 
in Eq.~(\ref{Dov}) is the correct formulation of the overlap operator for $\mu \ne 0$.

\section{Direct and iterative methods} \label{sec:directit}

A numerical implementation of the sign function using the spectral
definition \eqref{fA} is only possible for small
matrices, as a full diagonalization becomes too expensive as the matrix
grows.
As an alternative, matrix-based iterative algorithms for the
computation of the matrix sign function have been around for many
years \cite{Denman76,Kenney91,Kenney95,Higham97}.
Although these algorithms are much faster than the direct implementation
of the spectral definition for medium-sized problems, they still require
the storage and manipulation (i.e., inversions and/or multiplications)
of the entire matrix.
This is feasible for medium-sized matrices, but becomes too expensive
for the very large matrices occurring in typical lattice QCD
simulations.
E.g., even for an $8^{3} \times 8$ lattice, which is the
minimum lattice size required
for a physically relevant problem, the matrix dimension
is already $12 \cdot 8^{3} \cdot 8 \approx 50\,000$.
Even though these QCD matrices are sparse, the iterative procedure
computing the sign function fills the matrix as the iterations proceed,
and the method eventually becomes prohibitively expensive.

Therefore, another iterative method is required which does not produce
an  approximation to the full sign matrix itself, but rather produces
an approximation to 
the vector $y=\sign(A) x$, i.e., to the operation of the sign matrix on
an arbitrary vector. 
Many QCD applications only require the knowledge of this product for a
number of selected source vectors $x$.
For instance, some low-energy properties of QCD can be described by the
lowest-lying eigenvalues of the Dirac operator.
These eigenvalues can efficiently be found  by an iterative eigenvalue
solver like ARPACK \cite{arpack}, which only requires the computation of
matrix-vector multiplications. 
Analogously, the computation of the propagation of fermions can be well
approximated by inverting the Dirac operator on a selected number of
source vectors
$b_{k}$, i.e., the solution of the systems
$D_{\text{ov}} x = b_{k}$.
These inversions are also performed using iterative linear solvers
requiring only matrix-vector multiplications.  

Such iterative methods, mostly from the class of Krylov subspace methods,
are already extensively used for the solution of eigenvalue problems,
linear systems, and for function evaluations \cite{Vor88,Drus98} with
Hermitian matrices.
There, the ancestor of all methods is the Lanczos method, of which many
variants and improvements have been built over the years.
The Lanczos method makes use of short recurrences to build an
orthonormal basis in the Krylov subspace. 

Krylov subspace methods are also used for non-Hermitian matrices in the
context of eigenvalue problems (a popular example being the restarted
Arnoldi method of ARPACK), for the solution of linear systems, and even
for the evaluation of the exponential function
\cite{gallopoulos89parallel,hochbruck}.
The two most widely used methods are the Arnoldi method and the
two-sided Lanczos method.
In contrast to the Hermitian case, the Arnoldi method requires long
recurrences to construct an orthonormal basis for the Krylov subspace,
while the two-sided Lanczos method uses two short recurrence relations,
but at the cost of losing orthogonality. 

In the next section we present an Arnoldi-based method to compute a
generic function of a non-Hermitian matrix.  The application of the
two-sided Lanczos method to this problem will be investigated in a
separate publication.

\section{Arnoldi function approximation for a non-Hermitian matrix}%
\label{NumApp}

From the spectral definition \eqref{fA} it follows that $f(A)$ is
identical to some polynomial $P_{K-1}( A )$ of degree $K-1 < N$.
Indeed, the unique interpolation polynomial $P_{K-1}( z )$,
which interpolates $f(z)$ at the different
eigenvalues $\lambda_i$ of $A$, satisfies $f(A) = P_{K-1}(A)$,
as follows immediately  from Eq.~\eqref{fA}. If $A$ has
non-trivial Jordan blocks, this is still true, but interpolation 
is now in the Hermite sense, where at each eigenvalue $P_{K-1}$
interpolates $f$ and its
derivatives up to order one less than the size 
of the largest corresponding Jordan block, see \cite{HoJo}.

Hence, for an arbitrary vector $x$,
\be
y \equiv f(A) x = P_{K-1}( A )
x = \sum_{i=0}^{K-1} c_i A^i x \:.
\label{interp}
\ee

The idea is to construct an approximation to $y$ using a polynomial of
degree $k-1$ with $k \ll N$. 
One possibility is to construct a \emph{good} polynomial approximation
$P_{k-1}(A)$ such that $P_{k-1}(\lambda_i) \approx f(\lambda_i)$. 
This would yield a single polynomial approximation operating on any
vector $x$ to compute $f(A) x \approx P_{k-1}(A) x$.
One such example is the expansion in terms of Chebyshev polynomials up 
to degree $k - 1$.
Although Chebyshev polynomials are very close to the minimax polynomial
for a function approximation over an appropriate ellipse in the complex plane,
one can do better for matrix function approximations.
First of all, one only needs a good approximation to $f$ at the eigenvalues of
$A$, and secondly, one can use information about the source vector $x$ to
improve the polynomial approximation.
The vector $x$ can be decomposed using the (generalized) eigenvectors of $A$ as a
complete basis, and clearly some eigendirections will be more relevant
than others in the function approximation.
Using a fixed interpolation polynomial does not use any information
about the vector
$x$. Furthermore, the only feature of $A$ usually taken into account
by such polynomial approximations is the extent of its
spectrum.

Indeed, there exists a \emph{best} polynomial approximation
$\hat y = P_{k-1}(A) x$ of degree at most $k-1$, which is readily defined
as the orthogonal projection of $y$ on the Krylov subspace
${\cal K}_{k}(A,x) = \myspan(x, Ax, A^2x, \ldots, A^{k-1} x)$.
If $V_k=(v_1,\ldots,v_k)$ is an $N \times k$ matrix whose columns form an
orthonormal basis in $\K_{k}$, then $V_k V_k^\dagger$ is a projector on the
Krylov space, and the projection is $\hat y = V_k V_k^\dagger y$.

The operation of any polynomial of $A$ of degree smaller than $k$ on $x$ is
also an element of $\K_{k}$, and the projection $\hat y$ corresponds to
the polynomial which minimizes $\Delta = f(A) x - P_{k-1}(A) x$, as
$\Delta \perp {\cal K}_{k}(A,x)$. 
Clearly, this approximation explicitly takes into account information
about $A$ and $x$.

The problem, however, is that to find this \emph{best} polynomial
approximation of degree $k-1$ one already has to know the answer
$y = f(A) x$. 
Therefore, we need a method to approximate the projected vector
$\hat y$.  This can be done by one of the Krylov subspace methods
mentioned in Sec.~\ref{sec:directit}.  Here, 
we use the Arnoldi algorithm to construct an orthonormal basis for the
Krylov subspace $\K_k(A,x)$ using the long recurrence
\be
A V_k = V_k H_k + \beta_k v_{k+1} e_k^T \:,
\label{Arnoldi}
\ee
where $v_1=x/\beta$, $\beta=|x|$, $H_k$ is an upper Hessenberg matrix (upper
triangular + one subdiagonal), $\beta_k=H_{k+1,k}$, and $e_k$ is the
$k$-th basis vector in $\MathC^{k}$.
The projection $\hat y$ of $f(A) x$ on $\K_k(A,x)$ can be written as
\be
\hat y = V_k V_k^\dagger f(A) x \:.
\label{yproj}
\ee
Making use of $x = \beta v_1 = \beta V_k e_1$, Eq.~\eqref{yproj} becomes
\be
\hat y = \beta V_k V_k^\dagger f(A) V_k e_1 \:.
\label{yproj2}
\ee
From Eq.~\eqref{Arnoldi} it is easy to see that
\be
H_k = V_k^\dagger A V_k
\label{HVAV}
\ee
as $V_k^\dagger V_k = I_k$ and $v_{k+1} \perp V_k$.
Therefore it seems natural to introduce the approximation
\cite{gallopoulos89parallel}
\be
f(H_k) \approx V_k^\dagger f(A) V_k 
\label{fAapprox}
\ee
in Eq.~\eqref{yproj2}, which finally yields
\be
\hat y \approx \beta V_k f(H_k) e_1 \:.
\label{yproj3}
\ee
This expression is just a linear combination of the $k$ Arnoldi vectors
$v_i$, where the $k$ coefficients are given by $\beta$ times the first
column of $f(H_k)$.
Saad  \cite{saad:209} showed that the approximation \eqref{yproj3} corresponds to replacing
the polynomial interpolating $f$ at the eigenvalues of $A$ by the lower-degree
polynomial which interpolates $f$ at the eigenvalues of $H_k$, which
are also called the Ritz values of $A$ in the Krylov space.
The hope is that for $k$ not too large the approximation \eqref{yproj3}
will be a suitable approximation for $y$.
The approximation of $f(A) x$ by Eq.~\eqref{yproj3} replaces the
computation of $f(A)$ by that of $f(H_k)$, where $H_k$ is of much smaller size
than $A$.
The evaluation of (the first column of)
$f(H_k)$ can be implemented using a spectral decomposition, or another
suitable evaluation method \cite{Denman76,Kenney91,Kenney95,Higham97}.  

The long recurrences of the Arnoldi method make the method much slower
than the Lanczos method used for Hermitian matrices.
Nevertheless, our first results showed consistent convergence properties
when computing the matrix sign function: as the size of the Krylov space
increases, the method converges to within machine
accuracy.
More precisely, for the sign function the method shows a see-saw profile
corresponding to even-odd numbers of Krylov vectors, as was
previously noticed for the Hermitian case as well
\cite{vandenEshof:2002ms}. This even-odd pattern is related
  to the sign function being odd in its argument.
The see-saw effect is completely removed when using the
Harmonic Ritz values as described in
Ref.\ \cite{vandenEshof:2002ms} for Hermitian matrices and
extended to non-Hermitian matrices in Ref.\ \cite{hochbruck},
and convergence becomes smooth.  Alternatively, 
one can just as well restrict the calculations to even-sized Krylov subspaces.

Unfortunately, in the case of the sign function the Arnoldi method
described above has a very poor efficiency when some of the eigenvalues
are small.
In the case considered in Fig.~\ref{DeflArn} (see Sec.~\ref{Results}
below), the size of the Krylov 
space has to be taken very large ($k \approx N/2$) to reach accuracies
of the order of $10^{-8}$ (see the $m=0$ curve in the top pane of Fig.~\ref{DeflArn}).
A discussion and resolution of this problem are given in the next
section.

%%%%%%%%%%%%%%%%%%%%%%%%%%%%%%%%%%%%%%%%%%%%%%%%%%%%%%%%%%
%%%%%%%%%%%%%%%%%%%%%%%%%%%%%%%%%%%%%%%%%%%%%%%%%%%%%%%%%%

\section{Deflation}
\label{Deflation}

\subsection{Introduction}
\label{DeflIntro}

For Hermitian matrices, it is well known that the computation of the
sign function can be improved by deflating the eigenvalues smallest in
absolute value \cite{vandenEshof:2002ms}.  The reason why this is
crucial and specific to the sign function is the discontinuity of the
sign function at zero.  For non-Hermitian matrices, the situation is
analogous since the sign function now has a discontinuity along the
imaginary axis.
A necessary condition for the method described in
Sec.~\ref{NumApp} to be efficient is the ability to approximate $f$
well at the eigenvalues of $A$ by a low-order polynomial. 
If the gap between the eigenvalues of $A$ to the left and
  to the right of the imaginary axis is small,
no low-order polynomial will exist which will be
accurate enough for all eigenvalues.

The idea is to resolve this problem by treating these critical
eigenvalues exactly, and performing the Krylov subspace approximation
on a deflated space.

In the Hermitian case, deflation is straightforward.
The function $f(A)$ of a Hermitian matrix $A$ with eigenvalues
$\lambda_i$ and orthonormal eigenvectors $u_i$ ($i=1,\ldots,N$) can be
written as
\be
f(A) = \sum_{i=1}^N f(\lambda_i) u_i u_i^\dagger\:,
\ee
and its operation on an arbitrary vector $x$ as
\be
f(A) x = \sum_{i=1}^N f(\lambda_i)
( u_{i}^{\dagger} x ) u_i \:.
\label{fAxHerm}
\ee
If $m$ critical eigenvalues of $A$ and their corresponding eigenvectors
have been computed, one can split the vector space into two orthogonal
subspaces and write an arbitrary vector as $x=x_\parallel + x_\perp$,
where $x_\parallel=\sum_{i=1}^m ( u_{i}^{\dagger} x ) u_i$ and
$x_\perp = x - x_\parallel$.
Eq.~\eqref{fAxHerm} can then be rewritten as
\be
f(A) x = \sum_{i=1}^m f(\lambda_i) ( u_{i}^{\dagger} x )
u_i + f(A) x_\perp \:.
\label{fAxdefl}
\ee
The first term on the right-hand side of Eq.~\eqref{fAxdefl} can be
computed exactly, and the second term can be approximated by applying
the Arnoldi method of Sec.~\ref{NumApp} to $x_\perp$.
As the vector $x_\perp$ does not contain any contribution in the
eigenvector directions corresponding to the critical
eigenvalues, the polynomial approximation no longer needs to
interpolate $f$ at the eigenvalues closest to the function
discontinuity to approximate $f(A) x_\perp$ well.
Therefore, after deflation, lower-degree polynomials will yield the
required accuracy, and a smaller-sized Krylov subspace can be used in
the approximation.
In theory, the orthonormality of the eigenvectors guarantees
that the Krylov subspace will be perpendicular
to $x_\parallel$, but in practice numerical inaccuracies
could
require us to reorthogonalize the subspaces during the
construction of the Krylov subspace.

For non-Hermitian matrices the (generalized) eigenvectors are no longer orthonormal, 
and it is not immediately clear how to deflate a critical subspace. 
The matrix functions as defined in \eqref{fA} or \eqref{fAJordan} involve the 
inverse of the matrix of basis vectors, 
$U$, and no simple decomposition into orthogonal subspaces can be
performed.

In the remainder of this section, we will 
develop two alternative deflation schemes for the 
non-Hermitian case, using a composite subspace generated by adding a small number of critical eigenvectors to the Krylov subspace. 
This idea of an \emph{augmented Krylov
subspace method} has been used in the iterative solution of linear 
systems for some time, see, e.g., Ref.~\cite{Saad97}.
Since in computational practice one will never 
encounter non-trivial Jordan blocks, we assume in the following, for simplicity, that the matrix is diagonalizable.

%%%%%%%%%%%%%%%%%%%%%%%%%%%%%%%%%%%%%%%%%%%%%%%%%%%%%%%%%%%%%%%%%%%%%%%%%%%%%%%%

\subsection{Schur deflation}
\label{Schurdefl}

We construct the subspace $\Omega_m + {\cal K}_k(A,x)$, which is the sum of
the subspace $\Omega_m$ spanned by the eigenvectors corresponding
to $m$ critical eigenvalues of $A$ and the Krylov subspace ${\cal K}_k(A,x)$.
The aim is to make an approximation similar to that of
Eq.~\eqref{yproj3}, but to treat the contribution of the critical
eigenvalues to the sign function explicitly so that the
size of the Krylov subspace can be kept small. 

Assume that $m$ critical eigenvalues and right eigenvectors of $A$ are determined using an iterative eigenvalue solver like the one implemented in ARPACK.
From this one can easily construct $m$ Schur vectors and the corresponding $m \times m$ upper triangular matrix
$T_m$ satisfying
\be
A S_m = S_m T_m\:,
\label{pSd}
\ee
where $S_m=(s_1,\ldots, s_m)$ is the $N \times m$
matrix formed by the orthonormal Schur vectors and the diagonal
elements of $T_m$ are the eigenvalues corresponding to the Schur
vectors.
These Schur vectors form an orthonormal basis of the eigenvector
subspace $\Omega_m$, which is invariant under operation of $A$. 

After constructing the $m$-dimensional subspace $\Omega_m$ we run a
modified Arnoldi method
to construct an orthogonal basis of the composite subspace
$\Omega_{m} + {\cal K}_k(A,x)$.
That is, each Arnoldi vector is orthogonalized not only
against the previous ones, but also against the Schur vectors
$s_{i}$.
In analogy to \eqref{Arnoldi}, this process can be summarized as
\be
A \begin{pmatrix} S_{m} & V_{k} \end{pmatrix}
=
\begin{pmatrix} S_{m} & V_{k} \end{pmatrix}
\begin{pmatrix} T_{m} & S_{m}^{\dagger} A V_{k} \\
		 0 & H_{k} \end{pmatrix}
+ \beta_k v_{k+1} e_{m+k}^T \: .
\label{eq:modArnoldi}
\ee
Here, the columns $v_{1}, \ldots, v_{k}$ of $V_{k}$ form an
orthonormal basis of the space ${\cal K}_k^\perp(A,x)$, which
is the projection of the Krylov subspace $\K_k(A,x)$ onto
the orthogonal complement $\Omega^\perp$ of $\Omega_{m}$.
In particular, $v_1=x_\perp/\beta$, where
$x_\perp=( 1 - S_m S_m^\dagger ) x$ is the projection of
$x$ onto $\Omega^{\perp}$ and $\beta=|x_\perp|$.
Again, $H_k$ is an upper Hessenberg matrix.

Note that the orthogonality of $\K_k^\perp$ with respect to $\Omega_m$
has to be enforced
explicitly during the Arnoldi iterations, as the operation of $A$ on a
vector in the projected Krylov subspace $\K_k^\perp$ in general
gets a contribution belonging to $\Omega_m$, i.e., $\K_k^\perp$ is not
invariant under the operation of $A$.
This is a consequence of the non-orthogonality of the eigenvectors
of $A$.

Defining
\be
Q = \begin{pmatrix} S_m & V_{k} \end{pmatrix}
\ \mbox{and} \
H = \begin{pmatrix} T_m  & S_m^\dagger A V_k \\
0 & H_k \end{pmatrix} \: ,
\label{eq:newH}
\ee
$H$ satisfies a relation similar to Eq.~\eqref{HVAV}, namely
\be
H = Q^\dagger A Q \: ,
\label{eq:newHQHQ}
\ee
and the function approximation derived in Sec.~\ref{NumApp} can be used
here as well.
We briefly repeat the steps of Sec.~\ref{NumApp}.
The operation of the matrix function $f(A)$ on the vector $x$ can be
approximated by its projection on the composite subspace,
\be
f(A) x \approx Q Q^\dagger f(A) x \:,
\label{fAxSchur}
\ee
and because $x$ lies in the subspace,
\be
f(A) x \approx Q Q^\dagger f(A) Q Q^\dagger x \:.
\label{fAxNH}
\ee
As $H$
satisfies Eq.~\eqref{eq:newHQHQ},
we can introduce the same approximation as in Eq.~\eqref{fAapprox},
\be
f(H) \approx Q^\dagger f(A) Q \:,
\label{fH}
\ee
and substituting Eq.~\eqref{fH} in Eq.~\eqref{fAxNH} we construct
the function approximation
\be
f(A) x \approx Q f(H) Q^\dagger x \:.
\label{yproj4}
\ee

Because of the block structure \eqref{eq:newH} of the composite
Hessenberg matrix $H$, the matrix $f(H)$ can be written as 
\be
f(H) =
\begin{pmatrix}
f(T_m) & Y \\
0 & f(H_k)
\end{pmatrix} \:.
\label{eq:f_H}
\ee
The upper left corner is the function of the triangular Schur matrix
$T_m$ (which is again triangular), and the lower right corner is the
(dense) matrix function of the Arnoldi Hessenberg matrix $H_k$.
The upper right corner reflects the coupling between both subspaces
and is given by
the solution of the Sylvester equation
\be
T_m Y - Y H_k = f(T_m) X - X f(H_k)
\label{SylvEq}
\ee
with $X=S_m^\dagger A V_k$, which follows from the identity $f(H) H = H f(H)$.
Combining \eqref{yproj4} and \eqref{eq:f_H}, we obtain
\begin{eqnarray}
  f( A ) x
& \approx &
\begin{pmatrix}
   S_{m} & V_{k}
  \end{pmatrix}
  \begin{pmatrix}
   f( T_{m} ) & Y          \\
	  0      & f( H_{k} ) 
  \end{pmatrix}
  \begin{pmatrix}
	   S_{m}^{\dagger} \\
	   V_{k}^{\dagger}
  \end{pmatrix}
	x
  \notag
\\
& = &
 S_{m} f( T_{m} ) S_{m}^{\dagger} \: x
	+ \begin{pmatrix}
		 S_{m} & V_{k}
	  \end{pmatrix}
	  \begin{pmatrix}
		 Y \\
		 f( H_{k} )
	  \end{pmatrix}
	  V_{k}^{\dagger} \, x
\: .
  \label{eq:formula}
\end{eqnarray}
Note that $V_{k}^{\dagger} x = \beta e_{1}$.
Therefore only $f( T_{m} )$ and the first column of $Y$ and
$f(H_{k})$, i.e., the first $m + 1$ columns of $f( H )$, are
needed to evaluate \eqref{eq:formula}.
This information can be computed using the spectral
definition \eqref{fA} or some other suitable method
\cite{Denman76,Kenney91,Kenney95,Higham97}.
In the case of the sign function we chose to use Roberts' iterative method \cite{Rob80}
\be
S^{n+1} = \frac{1}{2} \left[S^n + (S^n)^{-1}\right]
\label{Roberts}
\ee
with $S^0 = A$, which converges quadratically to $\sign(A)$. 
Roberts' method is applied to compute $\sign(T_m)$ and $\sign(H_k)$. The matrix $Y$ is computed by solving the Sylvester equation \eqref{SylvEq} using the method described in Appendix~\ref{App:Sylvester}.

In the implementation one has to be careful to compute the deflated
eigenvectors to high enough
accuracy, as this will limit the overall accuracy of the function approximation.
When computing $f(A) x$ for several $x$, the partial Schur decomposition
\eqref{pSd} and the triangular matrix $f( T_{m} )$ need to be computed only once.
Only the modified Arnoldi method must be repeated for each new
vector $x$.

We summarize our algorithm for approximating $f(A) x$:
\begin{enumerate}
\item Determine the eigenvectors for $m$ critical eigenvalues of $A$ using ARPACK. Construct and store the corresponding Schur matrix $S_m$ and the upper triangular
   matrix $T_m = S_m^\dagger A S_m$.
   The columns of $S_m$ form an orthonormal basis of a subspace
   $\Omega_m$.
\item Compute the triangular matrix $f( T_{m} )$ using \eqref{Roberts}.
\item Compute $x_\perp = ( 1 - S_{m} S_{m}^{\dagger} ) x$.
\item Construct an orthonormal basis for the projected Krylov subspace
   $\K_k^\perp(A,x_\perp)$ using a modified Arnoldi method.
   The basis is constructed iteratively by orthogonalizing each new
   Krylov vector with respect to $\Omega_m$ and to all previous Arnoldi
   vectors, and is stored as columns of a matrix $V_k$.
   Also build the upper Hessenberg matrix $H = Q^\dagger A Q$ with
   $Q=(S_m,V_k)$.
\item Compute (column $m + 1$ of) $f(H)$ using Roberts' iterative method \eqref{Roberts} on $H_k$ and solving the Sylvester equation \eqref{SylvEq} for $Y$ as described in Appendix~\ref{App:Sylvester}.
\item Compute the approximation $f(A) x \approx Q f(H) Q^\dagger x$
   using
formula \eqref{eq:formula}.
\end{enumerate}
If $f(A) x$ has to be computed for several $x$,
only steps (iii)-(vi) need to be repeated for each vector $x$.

%%%%%%%%%%%%%%%%%%%%%%%%%%%%%%%%%%%%%%%%%%%%%%%%%%%%%%%%%%%%%%%%%%%%%%%%%%%%%%%%

\subsection{LR-deflation}
\label{LRdefl}

An alternative deflation in the same composite subspace $\Omega_m +{\cal K}_k(A,x)$ can be constructed using both the left and right eigenvectors corresponding to the critical eigenvalues. This deflation algorithm is a natural extension of the method described in Sec.~\ref{DeflIntro} from the Hermitian to the non-Hermitian case. A similar idea has been used for the iterative solution of linear systems \cite{MoZh04,Hasenfratz:2005tt}.

Assume that $m$ critical eigenvalues of $A$ have been computed together with their corresponding left and right eigenvectors by some iterative method like the one provided by ARPACK.
The right eigenvectors satisfy
\be
A R_m = R_m \Lambda_m
\label{Rev}
\ee
with $\Lambda_m$ the diagonal eigenvalue matrix for the $m$ critical eigenvalues and 
$R_m=(r_1,\ldots,r_m)$ the matrix of right eigenvectors (stored as
columns).  Similarly, the left eigenvectors obey 
\be
L_m^\dagger A = \Lambda_m L_m^\dagger\:,
\label{Lev}
\ee
where $L_m=(\ell_1,\ldots,\ell_m)$ is the matrix containing the left eigenvectors (also stored as columns).
For a non-Hermitian matrix, the left and right eigenvectors corresponding to different eigenvalues are orthogonal (for degenerate eigenvalues linear combinations of the eigenvectors can be formed such that this orthogonality property remains valid in general). Furthermore, if  the eigenvectors are normalized such that $\ell_i^\dagger r_i=1$, then clearly $L_m^\dagger R_m = I_m$, and $R_m L_m^\dagger$ is an oblique projector on the subspace $\Omega_m$ spanned by the right eigenvectors. 

Let us now decompose $x$ as
\be
x = x_{\parallel} + x_{\ominus} \:,
\label{xdec}
\ee
where $x_{\parallel} = R_m L_m^\dagger x$ is an oblique projection of $x$ on $\Omega_m$  and $x_{\ominus} = x-x_{\parallel}$.

Applying $f(A)$ to the decomposition \eqref{xdec} yields
\be
f(A) x =  f(A) R_m L_m^\dagger x + f(A) x_{\ominus} \:.
\label{fAxLR}
\ee
The first term on the right-hand side can be evaluated exactly using
\be
f(A) R_m L_m^\dagger x =  R_m f(\Lambda_m) L_m^\dagger x \:,
\label{fARLx}
\ee
which follows from Eq.~\eqref{Rev}, while
the second term can be approximated by applying the Arnoldi method described in Sec.~\ref{NumApp} to the vector $x_{\ominus}$. An orthonormal basis is constructed in the Krylov subspace ${\cal K}_k(A,x_{\ominus})$ using the recurrence 
\be
A V_k = V_k H_k + \beta_k v_{k+1} e_k^T \:,
\label{LRArnoldi}
\ee
where $v_1=x_{\ominus}/\beta$ and $\beta=|x_{\ominus}|$. By construction, $x_{\ominus}$ has no components along the $m$ critical (right) eigendirections, and successive operations of $A$ will yield no contributions along these directions either, hence $\K_k(A,x_\ominus)$ does not mix with $\Omega_m$.
In principle, numerical inaccuracies accumulated during the Arnoldi iterations might make it necessary to occasionally re-extract spurious components along the critical eigendirections. However, this turned out not to be necessary in our numerical calculations.

Applying the Arnoldi approximation \eqref{yproj3} to Eq.~\eqref{fAxLR} yields  the final approximation
\be
f(A) x \approx  R_m f(\Lambda_m) L_m^\dagger x  + \beta V_k f(H_k) e_1 \:.
\label{fAxLRArn}
\ee
Note that again only the first column of $f( H_{k} )$ is needed to evaluate Eq.~\eqref{fAxLRArn}. As before, $f(H_k)$ has to be computed using some suitable method.
The function $\sign(H_k)$ can be efficiently computed using Roberts' algorithm \eqref{Roberts}.

We summarize our algorithm for approximating $f(A) x$ in the LR-deflation scheme:
\begin{enumerate}
\item Determine the left and right eigenvectors for $m$ critical eigenvalues of $A$ using ARPACK.
   Store  the corresponding eigenvector matrices $L_m$ and $R_m$. 
\item Compute $f(\lambda_i)$ ($i=1,\ldots,m$) for the critical eigenvalues.
\item Compute $x_{\ominus} = ( 1 - R_{m} L_{m}^{\dagger} ) x$.
\item Construct an orthonormal basis for the Krylov subspace
   $\K_k(A,x_{\ominus})$ using the Arnoldi recurrence.
   The basis is constructed iteratively by orthogonalizing each new
   Krylov vector with respect to all previous Arnoldi
   vectors, and is stored as columns of a matrix $V_k$.
   Also build the upper Hessenberg matrix $H_k = V_k^\dagger A V_k$.
\item Compute (the first column of) $f(H_k)$ using Roberts' iterative method \eqref{Roberts}.
\item Compute the approximation to $f(A) x$ using Eq.~\eqref{fAxLRArn}.
\end{enumerate}
If $f(A) x$ has to be computed for several $x$,
only steps (iii)-(vi) need to be repeated for each vector $x$.

\subsection{Discussion}
\label{Sec:DeflDisc}

We briefly compare both deflation schemes. Although both schemes use the same composite subspace $\Omega_m + {\cal K}_k(A,x)$, 
they yield different function approximations resulting from a different subspace decomposition.

In the Schur deflation, the composite subspace is decomposed in a sum of two \textit{orthogonal} subspaces which are coupled by $A$, while in the LR-deflation 
the subspaces no longer mix, at the expense of losing orthogonality of
the two subspaces.

Accordingly, both schemes introduce different approximations to $f(A)
x$: the Schur deflation approximates the orthogonal projection of the
solution vector on the total composite space using Eq.~\eqref{fH},
while the LR-deflation first extracts the critical eigendirections and
only approximates the orthogonal projection of the residual vector on
the Krylov subspace using Eq.~\eqref{yproj3}. Therefore, in the Schur
deflation the components of $f(A) x$ along the Schur vectors become
more accurate as the Krylov subspace becomes larger, while in the
LR-deflation the components of $f(A)x$ along the critical
eigendirections can be computed exactly, independently of the size of
the Krylov subspace.  This is probably the reason for the observation
that, for fixed $m$ and $k$, the LR-deflation is slightly more
accurate than the Schur deflation, see Fig.~\ref{DeflArnproj} below.

Numerically, the LR-deflation has two advantages over the Schur
deflation. First, its Arnoldi method does not require
the deflated directions to be (obliquely) projected out of the Krylov
subspace because the subspaces do not
mix. Second, this absence of coupling between the subspaces means
that the LR-scheme has no analog of the Sylvester equation
\eqref{SylvEq}.

A downside of the LR-deflation is that both left and right eigenvectors need to be computed in the initialization phase, whereas the Schur deflation only needs the right eigenvectors.
Hence, the Schur deflation will have a shorter initialization phase, unless one needs to operate with both $f(A)$ and its adjoint $f(A)^\dagger$, in which case both sets of eigenvectors are needed anyways (for the latter, the roles of left and right eigenvectors are interchanged). 

In the next section, we will present numerical results obtained with our modified Arnoldi method.

%%%%%%%%%%%%%%%%%%%%%%%%%%%%%%%%%%%%%%%%%%%%%%%%%%%%%%%%%%%%%%%%%%%%
%%%%%%%%%%%%%%%%%%%%%%%%%%%%%%%%%%%%%%%%%%%%%%%%%%%%%%%%%%%%%%%%%%%%

\section{Results}
\label{Results}

We implemented the modified Arnoldi method proposed in the previous section to compute the sign function occurring in the overlap Dirac operator \eqref{Dov} of lattice QCD. 

\begin{figure}
\centering
\includegraphics[width=88mm]{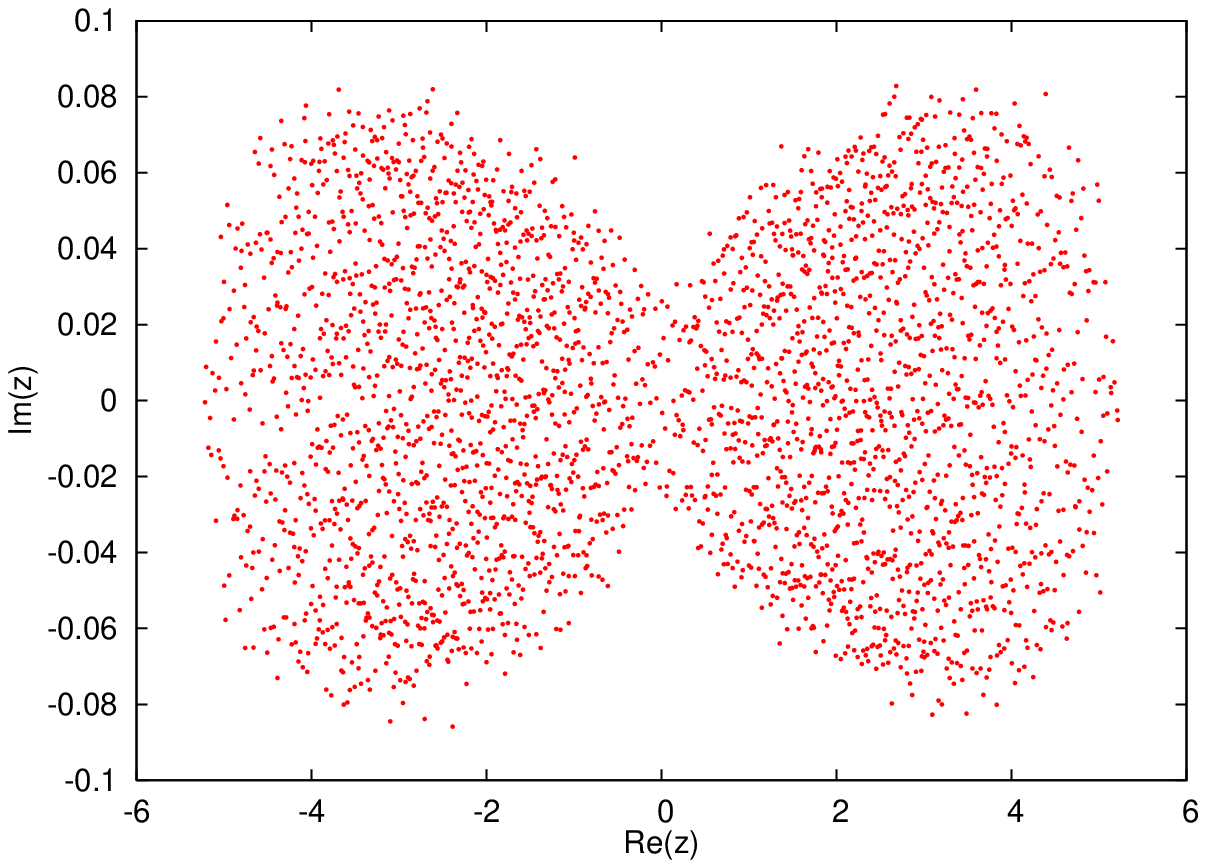}
\includegraphics[width=88mm]{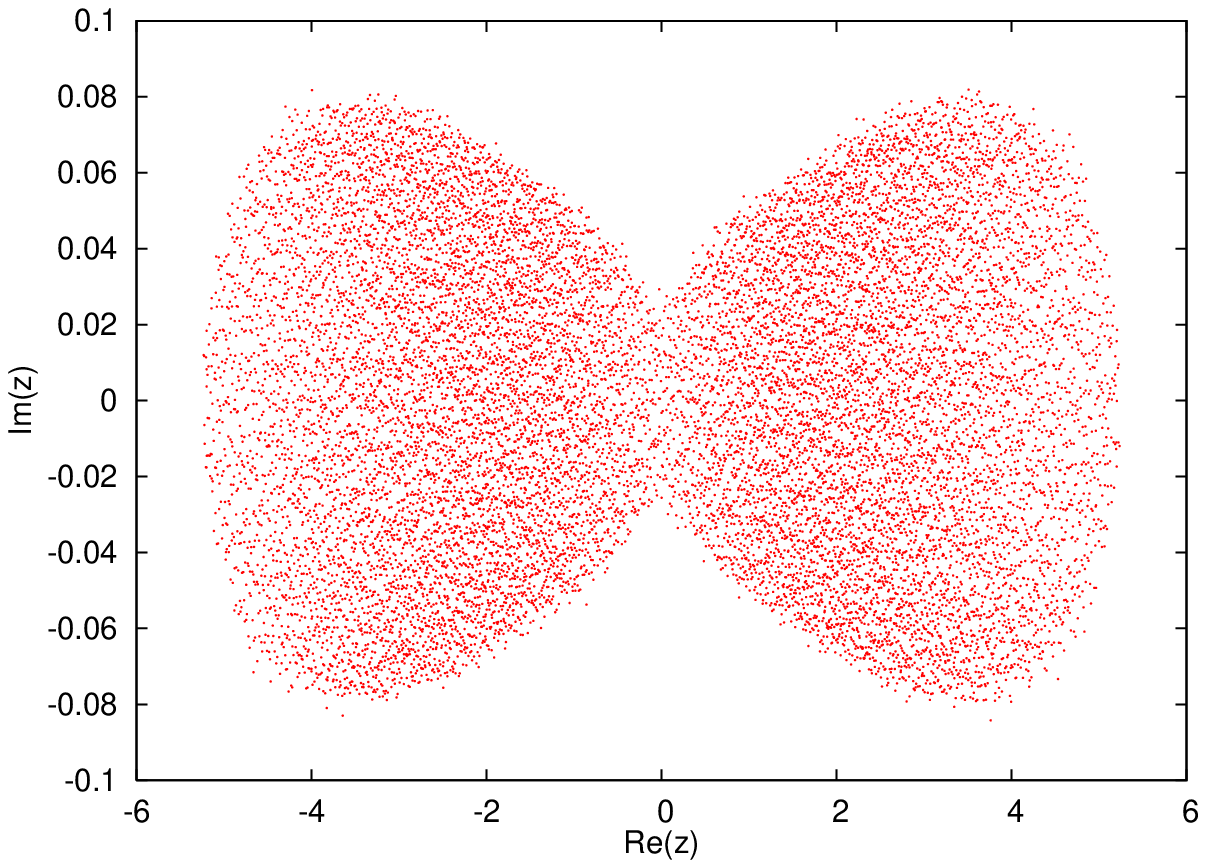}
\caption{Spectrum of $H_\text{w}(\mu)$ in Eq.~\eqref{Dov} for a $4^4$
  lattice (top pane) and a $6^4$ lattice (bottom pane), with $\mu=0.3$
  and $m_\text{w}=-2$.  Note the difference in scale between real and
  imaginary axes.  The gauge fields were generated using the Wilson
  plaquette action with gauge coupling $\beta_g=5.1$
  \cite{Bloch:2006cd}.}
\label{Fig:spectrumDw}
\end{figure}

First we discuss the critical eigenvalues of the $\gamma_5$-Wilson-Dirac operator $H_\text{w}(\mu)$, which are needed for deflation and have to be computed once for any given $SU(3)$ gauge configuration.
Deflation is needed because of the existence of eigenvalues close
to the imaginary axis.
In Fig.~\ref{Fig:spectrumDw} we show the spectrum of $H_\text{w}(\mu)$ for a
$4^4$ lattice and a $6^4$ lattice, using the same parameters as in
Ref.~\cite{Bloch:2006cd}, i.e., $m_\text{w}=-2$ and gauge coupling $\beta_g=5.1$. These complete spectra were
computed with LAPACK  \cite{lapack}. This is a very costly calculation, especially for the $6^4$ lattice, which was done for the sole purpose of this numerical investigation but cannot be performed routinely during production runs. 

Although the eigenvalues of interest for deflation in the case of the
sign function are those with smallest absolute real parts, we decided
to deflate the eigenvalues with smallest magnitude instead.
Numerically the latter are more easily determined, and both choices yield
almost identical deflations for the $\gamma_5$-Wilson operator at
nonzero chemical potential.  The reason for this is that, as long as
the chemical potential does not grow too large, the spectrum looks
like a very narrow bow-tie shaped strip along the real axis (see
Fig.~\ref{Fig:spectrumDw}), and the sets of eigenvalues with smallest
absolute real parts and smallest magnitudes will coincide.

\begin{figure}
\centering
\includegraphics[width=88mm]{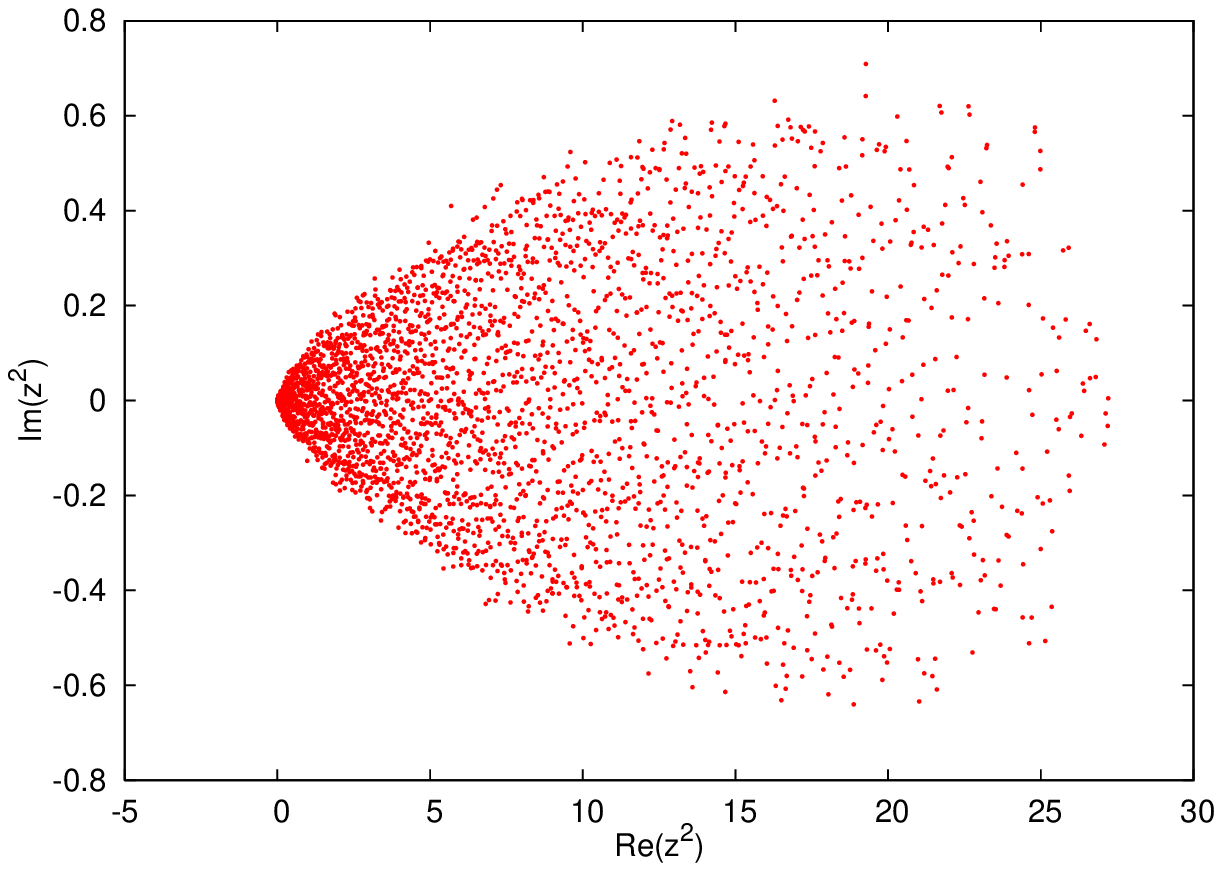}
\includegraphics[width=88mm]{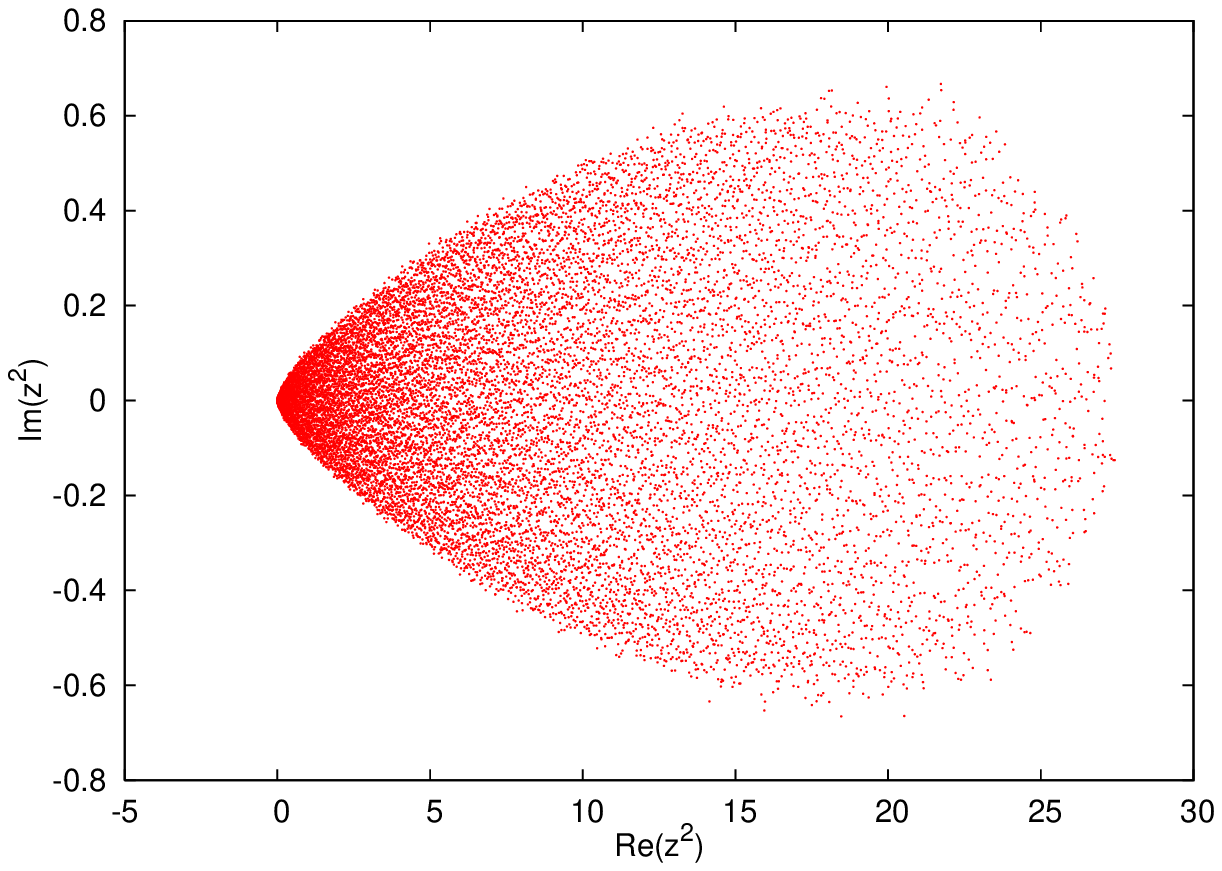}
\caption{Spectrum of $H_\text{w}^2(\mu)$ for a $4^4$ lattice (top pane) and a
  $6^4$ lattice (bottom pane), with $\mu=0.3$ and $\beta_g=5.1$.}
\label{Fig:spectrumDw2}
\end{figure}
In practice we compute the eigenvalues of $H_\text{w}$ with smallest magnitude with ARPACK. This package has an option to retrieve the eigenvalues with smallest magnitude without performing an explicit inversion, which would be very expensive in this case. However, the use of this option requires the eigenvalues to be near the boundary of the convex hull of the spectrum.
From Fig.~\ref{Fig:spectrumDw} it is 
clear that the eigenvalues closest to the origin are deep inside the interior of the 
convex hull and do not satisfy this requirement. 
Therefore we opted to compute the eigenvalues with smallest magnitude of the squared operator $H_\text{w}^2$. Clearly the eigenvalues with smallest magnitude will be the same for both operators, as $|\lambda^2| = |\lambda|^2$. The eigenvalues of $H_\text{w}^2$ are given by $z^2=x^2-y^2 + 2 i x y$, where $x$ and $y$ are the real and imaginary parts of the eigenvalues $z$ of $H_\text{w}$. The spectra of $H_\text{w}^2$ for the $4^4$ and $6^4$ lattices are shown in Fig.~\ref{Fig:spectrumDw2}, and clearly the eigenvalues with smallest magnitudes are now close to the boundary of the convex hull of the spectrum so that ARPACK can find them more easily. 
Since in this approach we square the matrix, there is the
potential danger that small eigenvalues get spoiled just by the additional numerical 
round-off introduced when applying $A$ twice. However, this should be noticeable only if
these eigenvalues are comparable in size to the round-off error. Our calculations are not affected by this problem.
 
Obviously there is a trade-off between the number of deflated eigenvalues and the size of the Krylov subspace. 
A useful piece of information in this context is the
ratio of the magnitude of the largest deflated eigenvalue over
the largest overall eigenvalue, which is given in Table~\ref{EVratio} for different numbers of deflated eigenvalues. 
A comparison of the values for both lattice sizes indicates that
the number of eigenvalues with a magnitude smaller than a given ratio
increases proportionally with the lattice volume. This is consistent
with a spectral density of the small eigenvalues proportional to the lattice volume. This property is also apparent in Figs.\ \ref{Fig:spectrumDw} and \ref{Fig:spectrumDw2}, as the contours enclosing the spectra remain unchanged when the volume is increased.
As a first guess we therefore expect that scaling $m$ with the volume should yield comparable convergence properties of the method for various lattice sizes.
\begin{table}[b]
\centering
\begin{tabular}{|c|c|c|}
\hline
$m$ & \multicolumn{2}{c|}{$\max |\lambda^\text{defl}|  / \max | \lambda^\text{all} |$} \\
\hline
& ~$4^4$ lattice~ & $6^4$ lattice
\\
\hline
2 & 0.0040 & 0.0016\\
4 & 0.0056 & 0.0026\\
8 & 0.0119 & 0.0035\\
16 & 0.0183 & 0.0052\\
32 & 0.0351 & 0.0084\\
64 & 0.0631 & 0.0155 \\
128 & --- & 0.0286 \\
\hline
\end{tabular}
\caption{Ratio of largest deflated eigenvalue
   over largest eigenvalue for various numbers of deflated eigenvalues
   $m$ for a $4^4$ and a $6^4$ lattice.}
\label{EVratio}
\end{table}

\begin{figure}
\centering
\includegraphics[width=88mm]{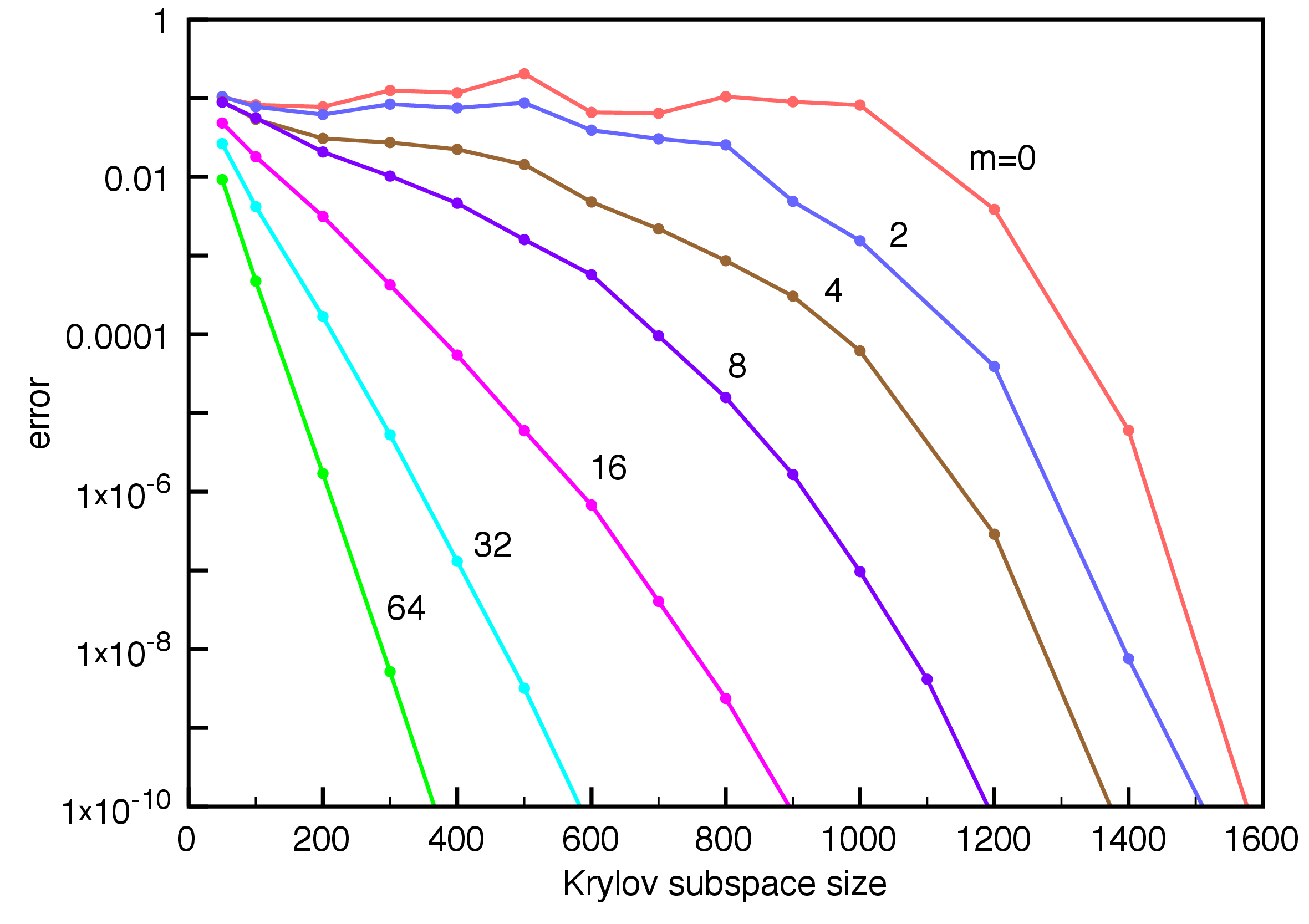}
\includegraphics[width=88mm]{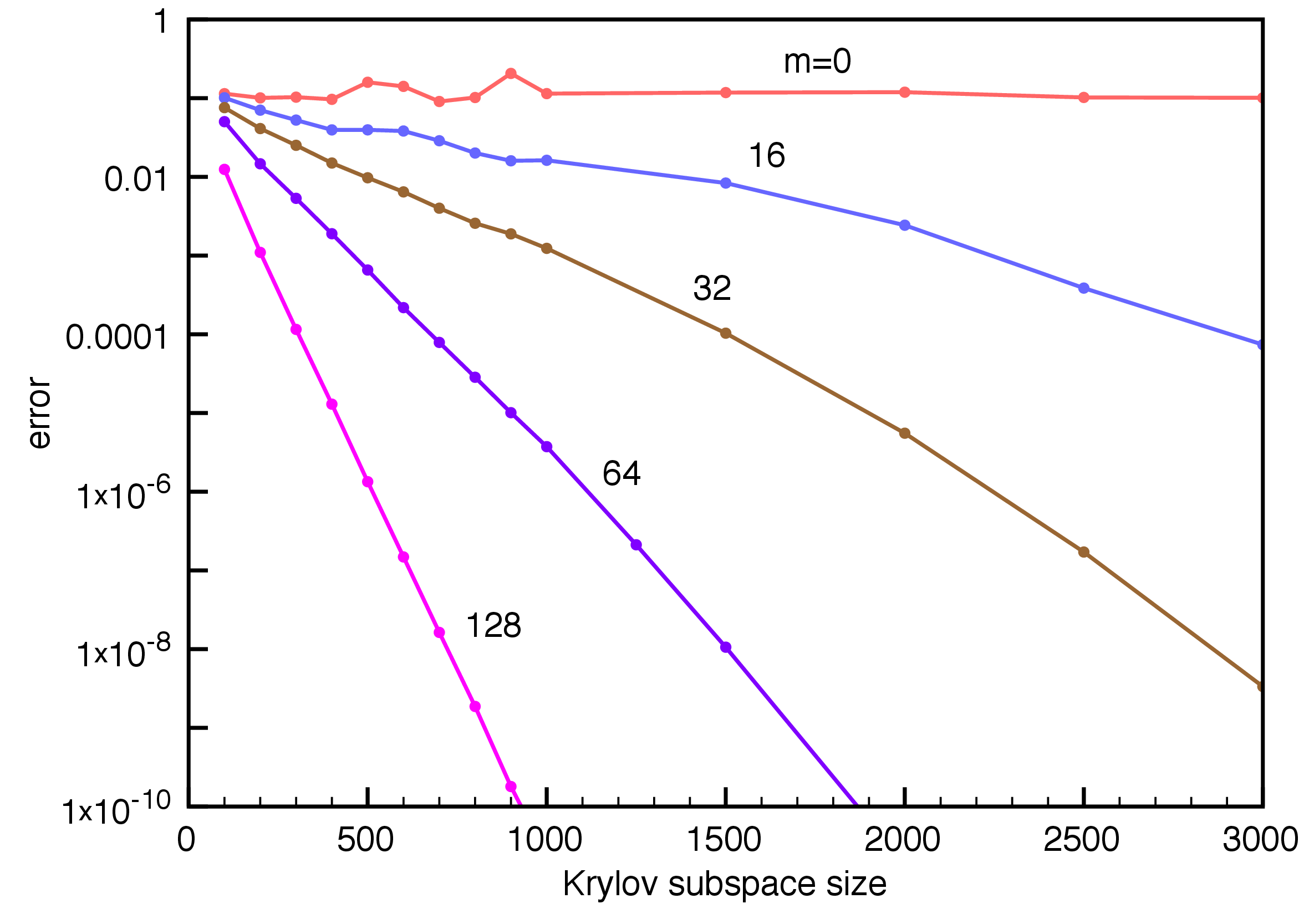}
\caption{\label{DeflArn}Accuracy of the modified Arnoldi method for
   $y=\sign(A) x$.
   The non-Hermitian matrix is $A=\gamma_5 D_\text{w}(\mu)$, where
   $D_\text{w}(\mu)$ is the Wilson-Dirac operator \eqref{Dw}
   at chemical potential $\mu=0.3$, and $x=(1,1,\ldots,1)$.
   \textit{Top pane}: $4^4$ lattice with $\dim(A)=3072$,
   \textit{bottom pane}: $6^4$ lattice with $\dim(A)=15552$. 
   The relative error $\epsilon=\| \tilde y - y \| / \| y \|
   $ is shown as a function of the Krylov subspace size $k$ for various
   numbers of deflated eigenvalues $m$ using the LR-deflation.
   In order to compute the error $\epsilon$, the exact solution $y$
   was first determined using the spectral definition \eqref{fA} and the
   full spectral decomposition computed with LAPACK.
   }
\end{figure}

The convergence of the method is illustrated in Fig.~\ref{DeflArn},
where the accuracy of the approximation is shown as a function of the
Krylov subspace size for two different lattice sizes.
The various curves correspond to different numbers of deflated
eigenvalues.
The results in the figure were computed using the LR-deflation scheme. The Schur deflation yields similar results.

Without deflation ($m=0$) the Krylov subspace method would be numerically unusable because of the need of large Krylov subspaces. 
Clearly, deflation highly improves the efficiency of the numerical method:
as more eigenvalues are deflated, smaller Krylov subspaces 
are sufficient to achieve a given accuracy. 

Furthermore, the deflation efficiency grows with increasing lattice volume. 
To reach an accuracy of $10^{-8}$ for the 
$4^4$ lattice with 25 ($\approx 0.0081 N$) deflated eigenvalues, one requires a Krylov subspace size $k \approx 570$ ($\approx 0.19 N$). However, to reach the same accuracy for the  
$6^4$ lattice with a comparable deflation of 128 ($\approx 0.0082 N$) critical eigenvalues, one only requires $k \approx 700$ ($\approx 0.045 N$).
Although the matrix size $N$ is more than 5 times larger for the $6^4$
lattice, the Krylov subspace only has to be expanded by a factor of 1.2 to achieve the same accuracy (when $m$ is scaled proportional to $N$ so that the ratio of Table~\ref{EVratio} remains approximately constant for both lattice sizes).

\begin{figure}
\centering
\includegraphics[width=88mm]{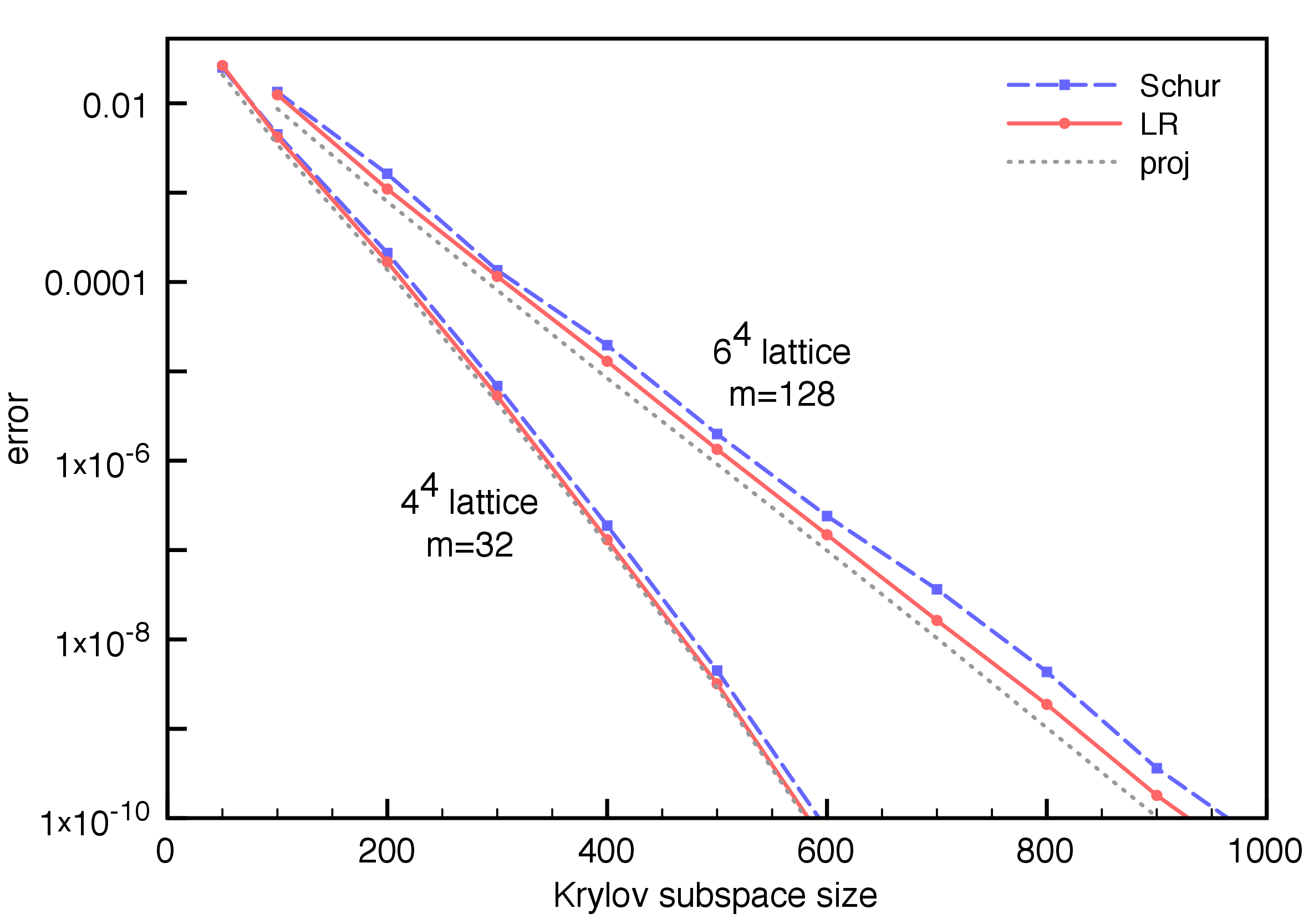}
\caption{\label{DeflArnproj}Comparison of the accuracies achieved with both deflation schemes and the exact projection of $y$ on the composite space $\Omega_m + {\cal K}_k(A,x)$. 
The relative error $\epsilon=\| \tilde y - y \| / \| y \|$ is shown as a function of the Krylov subspace size $k$. For both lattice sizes the accuracy of the LR-deflation is slightly better than that of the Schur deflation. Furthermore, the accuracy of the modified Arnoldi approximation is very close to the best possible approximation in the composite subspace.
}
\end{figure}

In Fig.~\ref{DeflArnproj} we compare the accuracy of the two deflation
schemes described in Sec.~\ref{Schurdefl} and \ref{LRdefl}. For an equal
number of deflated eigenvalues and equal Krylov subspace size, the
LR-deflation seems systematically slightly more accurate than the
Schur deflation.

To assess the quality of the modified Arnoldi approximation, it is interesting to compare the approximations \eqref{eq:formula} and \eqref{fAxLRArn}
for $f(A) x$ with the \emph{best approximation} in the composite
subspace, which corresponds to the orthogonal projection \eqref{fAxSchur} of $f(A) x$ on
$\Omega_m + {\cal K}_k(A,x)$,
\be
y_\text{proj} = Q Q^\dagger y = \sum_{i=1}^m
(s_{i}^{\dagger} y) s_i + \sum_{i=1}^k (v_i^{\dagger} y) v_i \:.
\ee
The relative accuracy of this projection is also shown in Fig.~\ref{DeflArnproj}.
It is encouraging to note that the modified Arnoldi approximation is
quite close to the exact projection $y_\text{proj}$.

\begin{table}[b]
\centering
\begin{tabular}{|c|c|c|c|}
\hline
\multicolumn{4}{|c|}{$4^4$ lattice, $m=32$}\\
\multicolumn{4}{|c|}{Schur deflation}\\
\hline\hline
\multicolumn{4}{|c|}{initialization time: 14.1 s}\\
\hline
$k$ & Arnoldi & $\sign(H)$ & ~total~ \\
\hline
100 &   0.18 & 0.03 & 0.23 \\
200 &   0.59 & 0.21 & 0.81 \\
300 &   1.22 & 0.52 & 1.75 \\                                                                      
400 &   2.05 & 1.08 & 3.16 \\
500 &   3.12 & 1.79 & 4.93 \\                                                                              
600 &   4.37 & 2.90 & 7.31 \\
700 &   5.88 & 4.57 & 10.49 \\
800 &   7.56 & 6.69 & 14.28 \\
900 &   9.50 & 9.38 & 18.92 \\
1000 &  11.63 & 12.68 & 24.36 \\
\hline
\end{tabular}
\hspace{2mm}
\begin{tabular}{|c|c|c|c|}
\hline
\multicolumn{4}{|c|}{$4^4$ lattice, $m=32$}\\
\multicolumn{4}{|c|}{LR-deflation}\\
\hline\hline
\multicolumn{4}{|c|}{initialization time: 27.5 s}\\
\hline
$k$ & Arnoldi & $\sign(H_k)$ & ~total~ \\
\hline
100 &    0.12 & 0.03 & 0.15    \\
200 &    0.45 & 0.20 & 0.66    \\
300 &    1.01 & 0.49 & 1.51    \\
400 &    1.77 & 1.02 & 2.82    \\
500 &    2.76 & 1.69 & 4.47    \\
600 &    3.94 & 2.77 & 6.74    \\
700 &    5.36 & 4.40 & 9.79    \\
800 &    6.96 & 6.44 & 13.44   \\
900 &    8.84 & 9.10 & 17.98   \\
1000 &   10.84 & 12.33 & 23.21 \\
\hline
\end{tabular}
\\[2mm]
\begin{tabular}{|c|c|c|c|}
\hline
\multicolumn{4}{|c|}{$6^4$ lattice, $m=128$}\\
\multicolumn{4}{|c|}{Schur deflation}\\
\hline\hline
\multicolumn{4}{|c|}{initialization time: 884 s}\\
\hline
$k$ & Arnoldi & $\sign(H)$ & ~total~ \\
\hline
100 &   2.03 & 0.05 & 2.13   \\
200 &   5.16 & 0.22 & 5.45   \\
300 &   9.27 & 0.56 & 9.91   \\
400 &   14.59 & 1.15 & 15.85 \\
500 &   20.95 & 2.09 & 23.17 \\
600 &   28.12 & 3.35 & 31.61 \\
700 &   36.81 & 5.17 & 42.15 \\
800 &   46.32 & 7.39 & 53.88 \\
900 &   56.83 & 10.37 & 67.39 \\
1000 &  68.29 & 13.88 & 82.39 \\
\hline                                                                                                                 
\end{tabular}
\hspace{2mm}                                                    
\begin{tabular}{|c|c|c|c|}
\hline
\multicolumn{4}{|c|}{$6^4$ lattice, $m=128$}\\
\multicolumn{4}{|c|}{LR-deflation}\\
\hline\hline
\multicolumn{4}{|c|}{initialization time: 1713 s}\\
\hline
$k$ & Arnoldi & $\sign(H_k)$ & ~total~ \\
\hline
100 &    0.66 & 0.03 & 0.75    \\            
200 &    2.39 & 0.15 & 2.62    \\
300 &    5.16 & 0.42 & 5.69    \\
400 &    9.01 & 0.94 & 10.06   \\
500 &    13.96 & 1.73 & 15.84  \\
600 &    20.03 & 2.80 & 22.98  \\
700 &    27.09 & 4.44 & 31.70  \\
800 &    35.09 & 6.49 & 41.78  \\                                                               
900 &    44.38 & 9.10 & 53.70  \\
1000 &   54.74 & 12.36 & 67.34 \\
\hline
\end{tabular}
\vspace*{2mm}
\caption{\label{CPUtime}CPU time (in seconds) for varying Krylov
   subspace size $k$. \textit{Top row:} $4^4$ lattice with $m=32$, \textit{bottom row:} $6^4$ lattice with $m=128$. \textit{Left panes:} Schur deflation, \textit{right panes:} LR-deflation. The time required by the initial calculation of the critical eigenvectors is given in the header of each block. The time needed to construct the Arnoldi basis in the Krylov subspace (column 2) is approximately proportional to 
$N k (k+2m)$ for the Schur deflation and $N k^2$ for the LR-deflation.
The time used by Roberts' method \eqref{Roberts} to compute the sign function of 
the Hessenberg matrix (column 3) is {\cal O}($k^3$).
The total time (column 4) also includes the evaluation of Eq.~\eqref{eq:formula} for the Schur deflation and Eq.~\eqref{fAxLRArn} for the LR deflation. These timings were measured on an Intel Core 2 Duo 2.33GHz computer using optimized ATLAS BLAS routines \cite{atlas-hp}.
   }
\end{table}

In Table~\ref{CPUtime} we show the CPU time used by the modified Arnoldi
method for the Schur and LR-deflation schemes. 
The times needed to construct the orthonormal basis in the Krylov
subspaces according to Eqs.~\eqref{eq:modArnoldi} and
\eqref{LRArnoldi} and to compute the sign function of the Arnoldi Hessenberg matrices are tabulated separately. 
The tabulated times were measured for an $m=32$ deflation for the
$4^4$ lattice and $m=128$ for the $6^4$ lattice. 

The larger CPU times required by the Schur deflation mainly reflect the additional orthogonalization of the Arnoldi vectors with respect to the Schur vectors. The time needed to compute the sign of the Hessenberg matrix is also slightly larger for the Schur deflation as it involves the additional solution of the Sylvester Equation \eqref{SylvEq}. 
For the same reasons, varying $m$ for a given lattice size will only significantly change the timings for the Schur deflation (this $m$-dependence is not shown in the table).

To summarize, the LR-deflation scheme has a somewhat better accuracy
and requires less CPU time per iteration than the Schur deflation. The
one advantage of the Schur deflation is that it only requires the
initial computation of the right eigenvectors, while the LR-deflation
requires the computation of both left and right eigenvectors. The time
needed to compute the critical eigenvectors of $H_\text{w}(\mu)$ is
given in the headers of the four blocks in Table~\ref{CPUtime}. The
choice of deflation scheme depends on the number of vectors $x$ for
which $\sign(H_\text{w}) x$ needs to be computed.  This will be the
topic of future work on nested iterative methods for non-Hermitian
matrices occurring during the inversion of the overlap operator.  Of
course, as mentioned in Sec.~\ref{Sec:DeflDisc}, if one needs to apply
both $\sign(H_\text{w})$ and its adjoint, then the LR-deflation will
be the better choice.

%%%%%%%%%%%%%%%%%%%%%%%%%%%%%%%%%%%%%%%%%%%%%%%%%%%%%%%%%%%%%%%%%%%%
%%%%%%%%%%%%%%%%%%%%%%%%%%%%%%%%%%%%%%%%%%%%%%%%%%%%%%%%%%%%%%%%%%%%

\section{Conclusion}

In this paper we have proposed an algorithm to approximate the action
of a function of a non-Hermitian matrix on an arbitrary vector, when
some of the eigenvalues of the matrix lie in a region of the complex
plane close to a discontinuity of the function.

The method approximates the solution vector in a composite subspace, i.e.,
a Krylov subspace augmented by the eigenvectors corresponding 
to a small number of critical eigenvalues.
In this composite subspace two deflation variants are presented based on different subspace decompositions: the Schur deflation uses two coupled orthogonal  subspaces, while the LR-deflation uses two decoupled but non-orthogonal subspaces.

The subspace decompositions are then
used to compute Arnoldi-based function approximations in which the contribution of the
critical eigenvalues is taken into account explicitly.
This deflation of critical eigenvalues allows for a smaller size of
the Krylov subspace and is crucial for the efficiency of the method.

For the sign function, deflation is particularly important because of
its discontinuity along the imaginary axis.
The method was applied to
the overlap Dirac operator of lattice QCD at nonzero chemical potential, 
where deflation was shown to
clearly enhance the efficiency of the method.
If the overlap Dirac operator has to be inverted using some
iterative method, each iteration will require the computation
of $\sign(H_\text{w}) x$ for some vector $x$.
In such a situation the cost for computing the critical
eigenvectors, which is done just once, is by far outbalanced
by the smaller costs for each evaluation of the sign function.
However, an important question that deserves further study is how the
optimal number $m$ of deflated eigenvectors depends on the volume and
how this influences the initialization time.
This question could become performance relevant for large volumes.

As mentioned above, our next steps include the application of the
two-sided Lanczos method to the problem of approximating $f(A)x$ for a
non-Hermitian matrix, and the investigation of nested iterative
methods for non-Hermitian matrices.  Work in these directions is in
progress. 

%%%%%%%%%%%%%%%%%%%%%%%%%%%%%%%%%%%%%%%%%%%%%%%%%%%%%%%%%%%%%%%%%%%%
%%%%%%%%%%%%%%%%%%%%%%%%%%%%%%%%%%%%%%%%%%%%%%%%%%%%%%%%%%%%%%%%%%%%

\appendix

\section{Sylvester equation}
\label{App:Sylvester}

In this appendix we describe a particularly simple algorithm to solve the special Sylvester equation
\be
T Y - Y H = C \:,
\label{GenSylvEq}
\ee
where $T$ is an $m \times m$ upper triangular matrix, $H$ is an $n \times n$ upper Hessenberg matrix, and the right-hand side $C$ and the unknown matrix $Y$ are $m \times n$ matrices. Classical methods to solve the Sylvester equation when $T$ and $H$ are full matrices are formulated in \cite{BaSt72,GoNaVL79}. 
For triangular $H$ and $T$ the Sylvester equation can easily be solved by direct substitution, see, e.g., \cite{Golub}.
In principle, this algorithm could also be applied to Eq.~\eqref{GenSylvEq} if the upper Hessenberg matrix $H$ is first transformed into triangular form using a Schur decomposition.  Here we present a more efficient approach, which can be regarded as a natural extension of the algorithm for the triangular Sylvester equation when one of the matrices is upper Hessenberg instead of triangular.
Blocking would also be possible (cf., e.g., \cite{Geijn}), but since
the solution of the Sylvester equation accounts only for a small
portion of the overall time we did not pursue this issue further.

Written out explicitly, the element $(i,j)$ of the matrix equation~\eqref{GenSylvEq} is
\be
\sum_{k=i}^m T_{ik} y_{kj} - \sum_{k=1}^{\max(j+1,n)} y_{ik} H_{kj} = c_{ij}
\label{SylvExpl}
\ee 
for $i=1,\ldots, m$ and $j=1, \ldots, n$.

This matrix equation can be solved row by row from bottom to top,
since Eq.~\eqref{SylvExpl} can be solved for row $i$ once rows $i+1, \ldots, m$ are known,
\be
T_{ii} y_{ij} - \sum_{k=1}^{\max(j+1,n)} y_{ik} H_{kj} = \tilde c_{ij}
\label{SolveRow}
\ee 
with $\tilde c_{ij}= c_{ij} - \sum_{k=i+1}^m T_{ik} y_{kj}$. 

Inside row $i$ one can solve for the element $y_{i,j+1}$ as a function of the elements to its left,
\be
y_{i,j+1} = -\frac{1}{H_{j+1,j}} \left[ \tilde c_{ij} - T_{ii} y_{ij} + \sum_{k=1}^j y_{ik} H_{kj}\right]
\label{SolveComp}
\ee 
for columns $j=1,\ldots, n-1$. From Eq.~\eqref{SolveComp} it follows that all elements of row $i$ can be written as 
\be
y_{ij} = a_j y_{i1} + b_j\:,
\label{xij}
\ee
where the coefficients $a_j$ and $b_j$ can be computed explicitly from the recurrence relations
\be
\begin{split}
a_{j+1} &= -\frac{1}{H_{j+1,j}} \left[- T_{ii} a_{j} + \sum_{k=1}^j a_{k} H_{kj}\right]\:, \\
b_{j+1} &= -\frac{1}{H_{j+1,j}} \left[ \tilde c_{ij} - T_{ii} b_{j} + \sum_{k=1}^j b_{k} H_{kj}\right]\:,
\end{split}
\label{abij}
\ee
starting from $a_1=1$, $b_1=0$. After substituting Eq.~\eqref{xij} with the known coefficients \eqref{abij}, element $(i,n)$ of Eq.~\eqref{SolveRow} can be solved for $y_{i1}$,
\be
y_{i1}  = \frac{\tilde c_{in} - T_{ii} b_n
+ \sum_{k=1}^n b_k H_{kn}}{T_{ii} a_n  - \sum_{k=1}^n a_k H_{kn}} \:.
\ee
Once $y_{i1}$ is known, all other elements $y_{ij}$ of row $i$ can be computed using Eq.~\eqref{xij} with coefficients \eqref{abij}.

\section*{Acknowledgments}
This work was supported in part by DFG grants FOR465-WE2332/4-2 and Fr755/15-1.
JB would like to thank Thomas Kaltenbrunner for useful discussions.

%%%%%%%%%%%%%%%%%%%%%%%%%%%%%%%%%%%%%%%%%%%%%%%%%%%%%%%%%%%%%%%%%%%%
%%%%%%%%%%%%%%%%%%%%%%%%%%%%%%%%%%%%%%%%%%%%%%%%%%%%%%%%%%%%%%%%%%%%

\bibliography{biblio}
\bibliographystyle{elsart-num}

\end{document}